\documentclass[11pt,a4paper]{article}

\usepackage[margin=1in]{geometry}
\usepackage{amsmath,amssymb,mathtools,bm}
\usepackage{amsthm}
\usepackage{authblk}
\usepackage{hyperref}
\usepackage{microtype}
\usepackage{setspace}
\usepackage{booktabs,tabularx}

\newcommand{\Var}{\operatorname{Var}}
\newcommand{\Cov}{\operatorname{Cov}}
\newcommand{\qs}{\mathrm{qs}}
\newcommand{\dd}{\mathrm{d}}
\newcommand{\mean}[1]{\left\langle #1\right\rangle}

\newcommand{\FR}{\mathrm{FR}}

\newtheorem{proposition}{Proposition}
\newtheorem{corollary}{Corollary}

\hypersetup{
    colorlinks=true,
    linkcolor=blue,
    citecolor=blue,
    urlcolor=blue
}

\title{Entropy Geometry and Condensation in Wealth Allocation}
\author[1]{Korak Biswas}
\affil[1]{GDQLabs Private Limited, Pune, India\\\texttt{korak@gdqlabs.com}}
\date{}

\begin{document}
\maketitle

\begin{abstract}
We develop a statistical framework for wealth allocation in which equilibrium-like statistics are derived from unbiased counting of admissible configurations rather than from postulated exchange rules. Each agent is characterized by a value--wealth convertibility function $V_i(w)$, whose local resolution determines the effective microscopic weight through a Jacobian relation. In a closed system, a canonical regular-sector distribution emerges from the microcanonical marginal and a reservoir expansion. The resulting partition sum yields a general critical-capacity criterion: when the regular sector reaches a finite absorption capacity, additional wealth condenses onto a small subset of agents. We extend the construction to open systems with variable wealth and agent number and to weakly driven quasistatic settings. The global wealth constraint then produces a constraint-induced evolution equation for the common parameter $\lambda(t)$, while simultaneous weak changes in wealth and value--wealth geometry give a unified first-order response. We further show that the regular-sector susceptibility $\chi_W=\sum_i\Var_i(w)$ is exactly the Fisher information of the joint canonical wealth family. Legendre duality gives the information-geometric identity $\dd s^2=\chi_W\dd\lambda^2=\chi_W^{-1}\dd W^2=-\mathcal{S}''(W)\dd W^2$. This separates condensation from metric divergence: a finite critical capacity requires $p>2$, whereas $\chi_W$ diverges only for $2<p\leq3$ and remains finite for $p>3$. The canonical boundary nevertheless lies at finite Fisher--Rao distance. A qualified Cram\'er--Rao duality and an open-system mixed-response relation follow, while contact-geometric, Airy-scaling, and stochastic-dynamical interpretations are identified explicitly as conjectures or future work. All established time-dependent and information-geometric results are quasistatic and do not determine a microscopic relaxation time.
\end{abstract}


\section{Introduction}

Understanding how wealth distributes across agents remains a central problem in complex systems and socioeconomic theory. A wide range of approaches have been developed to study wealth distributions, including stochastic growth models, agent-based exchange processes, and network-mediated interactions, many of which reproduce empirical features such as heavy-tailed distributions and scaling behavior \cite{dragulescu2000,yakovenko2009,bouchaud2000,solomon2001}. These approaches typically specify explicit mechanisms--rules of exchange, saving, growth, or institutional constraints--and then analyze the resulting steady states or long-time behavior. A complementary question is whether some large-scale features of wealth allocation can emerge even before specifying any particular mechanism, purely from the structure of the admissible economic configuration space.

At an abstract level, the problem can be viewed as distributing an extensive quantity among distinguishable agents, where each agent is associated with an internal multiplicity that depends on its local wealth. Closely related concentration phenomena are known in the econophysics literature, including wealth condensation in Pareto macroeconomies and exchange-driven models \cite{bouchaud2000,burda2002}, as well as condensation transitions in broader classes of occupancy and mass-transport models \cite{grosskinsky2003,evans2005,godreche2005,majumdar2005}. In most such settings, the functional weights controlling concentration are imposed through exchange rates, growth rules, or model-specific steady-state measures. The present work instead derives equilibrium-like allocations and concentration thresholds from static configuration counting, with the structural input encoded in a value-wealth mapping.

In this work, we introduce a framework in which wealth allocation is governed by unbiased counting of admissible configurations in a coarse-grained value-wealth space. The central object of the theory is a value-wealth convertibility function, which maps the wealth held by an agent to an effective scale associated with its internal productive differentiation. This function is treated as a primitive structural input, encoding how additional wealth expands, or fails to expand, the economically meaningful distinctions accessible at a given wealth level. Importantly, value is not identified with utility, payoff, or measured output; rather, it represents the resolution of productive configurations that are not tracked explicitly in the coarse-grained description.

Wealth is modeled as a non-negative extensive quantity distributed among distinguishable agents, with no upper bound on individual holdings. Within a given agent, wealth units are treated as indistinguishable at the level of description adopted here, while transfers between agents define distinct macrostates. The microscopic multiplicity associated with a given wealth allocation arises from internal configurations that are coarse-grained at the agent level and are not resolved explicitly. Entropy is defined directly from microstate counting \cite{huang1987}, without postulating any probabilistic form for the wealth distribution.

A central point of the present formulation is the distinction between two different uses of entropy. First, one may study local stationary allocations by extremizing the entropy of a specified macrostate vector under a fixed total wealth constraint. This gives a curvature condition for a stable interior saddle. Second, and more important for the stable condensation mechanism studied here, one may derive the marginal statistics of a regular sector by embedding the system in a reservoir. This second route yields a partition sum from counting alone. The existence of condensation is then decided by whether the regular-sector mean remains finite at the lower edge of the normalizable regime. No special tail form for the microscopic weight is assumed at the level of the general theory.

The framework isolates how the structural form of value--wealth convertibility shapes macroscopic wealth allocation. It provides a minimal description of stable regular phases and capacity-driven concentration, all derived from configuration counting without assuming explicit behavioral rules or dynamical growth models. Instability-driven concentration can also occur when no interior or regular description is meaningful, but the main focus of this paper is the stable capacity-driven route, in which the regular sector remains well defined while excess wealth forms a condensate. We additionally construct a controlled weak-time-dependence extension in which total wealth and value--wealth geometry vary on a slow timescale. This yields an instantaneous condensation threshold, a constraint-induced evolution law for the common parameter $\lambda$, and a critical wealth susceptibility, while leaving explicit transaction dynamics outside the model.

A further aim is to expose the information-geometric structure already implicit in this susceptibility. The joint canonical wealth family is an exponential family whose Fisher information is exactly $\chi_W$. Expressing the same metric in the Legendre-conjugate variables $\lambda$ and $W$ links the response coefficient to the curvature of the constrained entropy. This produces an information-geometric refinement of the critical classification: finite-capacity condensation occurs for $p>2$, but Fisher-metric divergence occurs only in the infinite-variance sector $2<p\leq3$. We also derive a qualified inference duality and a mixed-response identity for the open-system variables, while keeping contact-geometric and dynamical extensions explicitly separate from the established quasistatic results.

\section{Model and Assumptions}

We consider a closed system of $N$ distinguishable agents labeled by $i=1,\ldots,N$. Each agent holds a non-negative integer amount of wealth $w_i$, interpreted as a number of wealth quanta. Within a given agent, wealth quanta are treated as indistinguishable, while transfers between agents define distinct macrostates. No upper bound is imposed on individual wealth. The total wealth is conserved,
\begin{equation}
    \sum_{i=1}^{N} w_i = W,
    \label{eq:constraint}
\end{equation}
which constitutes the only global constraint.

\subsection{Macrostates and microstates}

A macrostate of the system is defined by the wealth-occupancy vector
\begin{equation}
    \bm{w}=(w_1,w_2,\ldots,w_N).
\end{equation}
Configurations that differ only by internal rearrangements of wealth within a given agent are regarded as microscopically distinct but macroscopically equivalent. By contrast, transferring a wealth quantum between agents changes the macrostate. This definition induces a coarse-graining in which microstates are grouped according to agent-level wealth occupancy.

For each agent $i$, we associate a number $\Omega_i(w)$ of accessible microstates with holding wealth $w$. These microstates represent distinct internal configurations that are coarse-grained at the agent level and are not resolved explicitly. At the level of description adopted here, all such configurations compatible with the same wealth are treated as accessible and contribute equally to the statistical weight.

With this convention, the total number of microstates compatible with a macrostate $\bm{w}$ factorizes as
\begin{equation}
    \Omega(\bm{w})=\prod_{i=1}^{N}\Omega_i(w_i).
    \label{eq:factorized}
\end{equation}
Agents are distinguishable and wealth transfers between agents define distinct macrostates. As a result, no additional combinatorial factors couple different agents, and the total multiplicity factorizes as in Eq.~\eqref{eq:factorized}.

\subsection{Value-wealth convertibility}

Each agent $i$ is characterized by a value-wealth convertibility function $V_i(w)$, which maps wealth $w$ to an effective value scale associated with the agent's internal productive capacity. The function $V_i(w)$ is introduced as a primitive structural property of the agent and is not defined through entropy, utility, payoff, or state counting. We treat wealth as a continuous variable when taking derivatives, corresponding to a coarse-grained limit of large wealth quanta. Unless otherwise stated, we assume
\begin{equation}
    V_i'(w)>0, \qquad V_i''(w)<0,
    \label{eq:concavity}
\end{equation}
reflecting monotonic increase with diminishing marginal convertibility. No specific functional form is imposed, and heterogeneity across agents is allowed.

\subsection{Microstate density and Jacobian postulate}

To relate internal structure to value-wealth convertibility, we postulate that the local density of accessible microstates in wealth space is governed by the resolution of the mapping between wealth and value.

\medskip
\noindent\textbf{Jacobian postulate.} The number of accessible microstates associated with holding wealth $w$ is proportional to the inverse local resolution of the value-wealth mapping,
\begin{equation}
    \Omega_i(w)\propto \left|\frac{dV_i}{dw}\right|^{-1}.
    \label{eq:jacobian}
\end{equation}
This expresses the idea that when marginal value resolution is low, many microscopically distinct internal configurations correspond to the same macroscopic wealth level. Multiplicative constants independent of $w$ are irrelevant for entropy differences and normalization ratios.

\subsection{Entropy}

The entropy associated with a macrostate $\bm{w}$ is defined by logarithmic microstate counting,
\begin{equation}
    S(\bm{w})=\ln \Omega(\bm{w})=\sum_{i=1}^{N}\ln \Omega_i(w_i).
\end{equation}
Using the Jacobian postulate, the entropy may be written, up to an additive constant, as
\begin{equation}
    S(\bm{w})=-\sum_{i=1}^{N}\ln V_i'(w_i)+\mathrm{const}.
    \label{eq:entropy}
\end{equation}
It is useful to define the single-agent entropy weight
\begin{equation}
    h_i(w)\equiv \ln \Omega_i(w)=-\ln V_i'(w)+\mathrm{const}.
    \label{eq:hdef}
\end{equation}

\section{Local Entropy Geometry and Stationarity}

The equilibrium wealth allocation of a specified macrostate vector may be studied by extremizing the entropy $S(\bm{w})$ subject to the conservation of total wealth. Introducing a Lagrange multiplier $\lambda$ to enforce Eq.~\eqref{eq:constraint}, we consider
\begin{equation}
    \mathcal{L}=S(\bm{w})-\lambda\left(\sum_{i=1}^{N}w_i-W\right).
\end{equation}
Using Eq.~\eqref{eq:entropy}, the stationarity condition is
\begin{equation}
    -\frac{V_i''(w_i)}{V_i'(w_i)}=\lambda,
    \label{eq:stationarity}
\end{equation}
for all agents whose wealth lies in the interior of the feasible domain. Equivalently,
\begin{equation}
    h_i'(w_i)=\lambda.
\end{equation}
The Lagrange multiplier $\lambda$ is conjugate to wealth and is determined by the global constraint when an interior solution exists.

\subsection{Interior stability}

The stationarity condition does not by itself guarantee a stable entropy maximum. Stability of the constrained entropy maximum requires the second variation to be negative on the constraint manifold. Since the entropy is additive and the constraint is linear, a sufficient local condition for interior stability is
\begin{equation}
    h_i''(w)<0.
    \label{eq:stableh}
\end{equation}
In terms of the convertibility function, this condition is
\begin{equation}
    \frac{d}{dw}\left[-\frac{V_i''(w)}{V_i'(w)}\right]<0.
    \label{eq:curvature_decreasing}
\end{equation}
Thus, the curvature ratio must decrease with wealth. The concavity condition $V_i''(w)<0$ is not sufficient by itself to ensure Eq.~\eqref{eq:curvature_decreasing}; it only ensures that the curvature ratio is positive.

This local saddle analysis is useful because it identifies when a deterministic interior allocation is stable. However, finite-capacity condensation in an unbounded wealth domain is more naturally characterized by the counting-derived marginal distribution derived below. In that formulation, the regular sector can remain statistically stable up to a critical value of $\lambda$, while excess wealth is carried by a condensate. The criterion for this stable condensation is not Eq.~\eqref{eq:curvature_decreasing} alone, but the finite critical capacity of the regular-sector partition sum.

\subsection{Curvature-ratio regimes}

The preceding local criterion should be interpreted together with the normalizability criterion developed in Sec.~\ref{sec:marginal}. In an unbounded wealth domain, the behavior of the curvature ratio
\begin{equation}
    r_i(w)\equiv -\frac{V_i''(w)}{V_i'(w)}=h_i'(w)
\end{equation}
separates three qualitatively different regimes.

First, if $r_i(w)$ is decreasing, then $h_i(w)$ is concave and the fixed-wealth entropy admits a stable interior saddle when the stationarity equation can be solved. However, this regime does not by itself produce a finite critical capacity. Since $r_i(w)$ approaches its lower edge from above, the critical weight at that edge is not sufficiently suppressed to give a finite regular-sector capacity on an unbounded wealth domain. This gives a stable regular sector without capacity-driven condensation.

Second, if $r_i(w)$ increases toward a finite limiting value $\lambda_{c,i}$, the local saddle criterion is not satisfied, but the counting-derived canonical regular sector may still be well defined for $\lambda>\lambda_{c,i}$. In that case the decisive question is whether the critical distribution at $\lambda=\lambda_{c,i}$ has a finite first moment. If it does, the regular sector has finite capacity and capacity-driven condensation occurs for $W>W_c$.

Third, if $r_i(w)$ grows without bound, then $h_i(w)$ grows faster than linearly. The exponential damping factor $e^{-\lambda w}$ cannot normalize the regular-sector distribution for any finite $\lambda$, and the system is driven directly toward boundary-dominated, instability-driven concentration.

Thus, the curvature-ratio test and the partition-sum test diagnose different aspects of the theory. The former identifies local interior-saddle stability. The latter decides whether a regular statistical sector exists and whether that sector has finite absorptive capacity.

\subsection{Interpretation of the intensive parameter}

The intensive parameter $\lambda$ has a direct structural interpretation. From Eq.~\eqref{eq:stationarity}, $\lambda$ quantifies the local rate at which marginal value convertibility saturates with increasing wealth. Large values of $\lambda$ correspond to rapid saturation of marginal convertibility, while smaller values indicate a greater ability of the regular sector to accommodate wealth. In the marginal formulation, the same parameter appears as the exponential weight generated by the reservoir entropy expansion.

\section{Stable Regular Sector and Capacity-Driven Condensation}

We now formulate the stable condensation mechanism of interest. The regular sector is defined as the part of the system whose wealth statistics are described by a normalizable single-agent distribution derived from microcanonical counting. It is called regular because typical agents remain microscopic relative to the total wealth, so that the reservoir expansion is valid. Condensation occurs only when this regular sector has a finite critical capacity.

\subsection{Regular-sector capacity}

For each agent $i$, define the counting-derived partition sum
\begin{equation}
    Z_i(\lambda)=\sum_{w=0}^{\infty}\Omega_i(w)e^{-\lambda w}.
    \label{eq:partition_intro}
\end{equation}
This object is not an additional thermodynamic postulate. It is derived in Sec.~\ref{sec:marginal} from the microcanonical marginal by expanding the entropy of the remaining agents. The microscopic weight $\Omega_i(w)$ is the same weight obtained from state counting and, under the Jacobian postulate, is proportional to $1/V_i'(w)$.

Let
\begin{equation}
    \lambda_{c,i}=\inf\{\lambda: Z_i(\lambda)<\infty\}
\end{equation}
be the lower edge of the normalizable regime for agent $i$, and let
\begin{equation}
    \lambda_c=\max_i \lambda_{c,i}.
    \label{eq:lambdac}
\end{equation}
For $\lambda>\lambda_c$, all regular-sector marginals are normalizable. The mean wealth of agent $i$ in the regular sector is
\begin{equation}
    \langle w\rangle_i(\lambda)=-\frac{\partial}{\partial\lambda}\ln Z_i(\lambda).
    \label{eq:mean_intro}
\end{equation}
The regular-sector capacity is defined by the critical limit
\begin{equation}
    W_c=\sum_{i=1}^{N}\langle w\rangle_{i,c},
    \qquad
    \langle w\rangle_{i,c}=\lim_{\lambda\downarrow\lambda_c}\langle w\rangle_i(\lambda),
    \label{eq:capacity}
\end{equation}
provided the limit exists and is finite.

Equation~\eqref{eq:capacity} is the central criterion. Concavity or diminishing marginal convertibility does not automatically imply finite capacity. Finite capacity is a structural property of the counted microscopic weights themselves. Once $\Omega_i(w)$ is specified by the value-wealth geometry, the question is whether the corresponding partition sum has a finite critical mean. No particular asymptotic form of $\Omega_i(w)$ is required at the level of the general theory.

\subsection{Capacity-driven condensation}

If $W_c$ is finite, the regular sector can accommodate at most $W_c$ units of wealth in the critical limit. For total wealth
\begin{equation}
    W\leq W_c,
\end{equation}
the conservation constraint can be satisfied by a value $\lambda>\lambda_c$ such that
\begin{equation}
    W=\sum_{i=1}^{N}\langle w\rangle_i(\lambda).
\end{equation}
For
\begin{equation}
    W>W_c,
\end{equation}
no normalizable regular-sector distribution with $\lambda>\lambda_c$ can absorb the imposed wealth. The excess wealth
\begin{equation}
    W_{\rm cond}=W-W_c
\end{equation}
therefore accumulates on a small subset of agents. This is a capacity-driven condensation transition.

The transition is stable in the following sense: the regular sector does not become ill defined or dynamically unstable at the transition. Instead, $\lambda$ becomes pinned at $\lambda_c$, the regular sector remains at its critical distribution, and additional wealth is carried by the condensate. Thus, the condensate arises from finite absorptive capacity, not from the loss of a regular statistical description.

\subsection{Relation to local entropy extremization}

The local saddle condition of Sec.~3 and the capacity criterion above are related but not identical. The saddle condition asks whether a particular interior wealth vector is a local maximum of $S(\bm{w})$. The capacity criterion asks whether the ensemble of regular configurations has a finite critical mean. In an unbounded wealth domain, the second question is the appropriate one for stable condensation, because the condensate is a collective large-deviation phenomenon: typical agents remain distributed according to the critical regular-sector law, while a small subset carries the excess wealth.

This distinction avoids imposing any special asymptotic form by hand. The theory first derives the statistical weights from counting; only after this derivation does one examine whether $Z_i(\lambda)$ has a finite critical mean. Different value-wealth geometries may therefore lead to no condensation, finite-capacity condensation, or instability-dominated concentration.

\section{Microcanonical Marginal and Emergent Canonical Form}\label{sec:marginal}

In this section we derive the marginal wealth statistics of a single agent directly from microcanonical state counting. This construction introduces no dynamical assumptions and makes explicit how an exponential weight emerges from the global wealth constraint.

\subsection{Microcanonical marginal}

Fix an agent $i$ and consider the probability that it holds wealth $w$ under the microcanonical constraint $\sum_{j=1}^{N}w_j=W$. The total number of microstates compatible with the event $w_i=w$ factorizes as the product of the internal multiplicity of agent $i$ and that of the remaining $N-1$ agents:
\begin{equation}
    \Omega(W;w_i=w)=\Omega_i(w)\Omega_{\rm res}(W-w),
\end{equation}
where $\Omega_{\rm res}(W-w)$ denotes the number of microstates accessible to the remaining agents under total wealth $W-w$. The corresponding microcanonical marginal is
\begin{equation}
    P_i(w)=\frac{\Omega_i(w)\Omega_{\rm res}(W-w)}{\Omega(W)}.
\end{equation}
Defining the reservoir entropy $S_{\rm res}(x)=\ln\Omega_{\rm res}(x)$, we may write
\begin{equation}
    P_i(w)\propto \Omega_i(w)e^{S_{\rm res}(W-w)}.
\end{equation}

\subsection{Reservoir expansion and canonical factor}

For agents in the regular sector, typical wealth fluctuations satisfy $w\ll W$, and the reservoir entropy may be expanded about $W$:
\begin{equation}
    S_{\rm res}(W-w)=S_{\rm res}(W)-\lambda w+\frac{\kappa}{2}w^2+\cdots,
\end{equation}
where
\begin{equation}
    \lambda\equiv \left.\frac{\partial S_{\rm res}}{\partial W}\right|_W,
    \qquad
    \kappa\equiv \left.\frac{\partial^2 S_{\rm res}}{\partial W^2}\right|_W.
\end{equation}
Retaining the leading term gives
\begin{equation}
    e^{S_{\rm res}(W-w)}\simeq e^{S_{\rm res}(W)}e^{-\lambda w},
\end{equation}
and therefore
\begin{equation}
    P_i(w)\propto \Omega_i(w)e^{-\lambda w}.
    \label{eq:canonical_weight}
\end{equation}
The exponential factor has therefore emerged from the wealth constraint through the reservoir entropy expansion. It has not been assumed as a Boltzmann form.

\subsection{Incorporation of value-wealth geometry}

Using the Jacobian postulate, $\Omega_i(w)\propto |dV_i/dw|^{-1}$, the marginal distribution takes the explicit form
\begin{equation}
    P_i(w)=\frac{1}{Z_i(\lambda)}\frac{e^{-\lambda w}}{V_i'(w)},
    \label{eq:explicit_marginal}
\end{equation}
where
\begin{equation}
    Z_i(\lambda)\propto \sum_{w=0}^{\infty}\frac{e^{-\lambda w}}{V_i'(w)}.
    \label{eq:Zexplicit}
\end{equation}
Thus, the exponential factor encodes the global conservation law, while deviations from a pure exponential are controlled by the value-wealth geometry through $V_i'(w)$.

\subsection{Normalizability and critical capacity}

The normalization of $P_i(w)$ requires that
\begin{equation}
    Z_i(\lambda)=\sum_{w=0}^{\infty}\Omega_i(w)e^{-\lambda w}
    \label{eq:Zi_general}
\end{equation}
be finite. The lower edge of the normalizable regime is
\begin{equation}
    \lambda_{c,i}=\inf\{\lambda:Z_i(\lambda)<\infty\}.
\end{equation}
For a heterogeneous system, the common regular-sector parameter must satisfy $\lambda>\lambda_c$, with $\lambda_c$ defined in Eq.~\eqref{eq:lambdac}. The mean wealth in the regular sector is
\begin{equation}
    \langle w\rangle_i(\lambda)
    =\frac{\sum_{w=0}^{\infty}w\,\Omega_i(w)e^{-\lambda w}}{\sum_{w=0}^{\infty}\Omega_i(w)e^{-\lambda w}}
    =-\frac{\partial}{\partial\lambda}\ln Z_i(\lambda).
    \label{eq:mean_general}
\end{equation}
A capacity-driven condensation transition is possible only if
\begin{equation}
    \langle w\rangle_{i,c}
    =\lim_{\lambda\downarrow\lambda_c}\langle w\rangle_i(\lambda)
    \label{eq:critical_mean}
\end{equation}
is finite for the regular sector, so that
\begin{equation}
    W_c=\sum_{i=1}^{N}\langle w\rangle_{i,c}<\infty.
\end{equation}
For $W\leq W_c$, the constraint is satisfied by a suitable $\lambda>\lambda_c$. For $W>W_c$, no normalizable regular distribution can accommodate the imposed wealth, and the excess $W-W_c$ must accumulate into a condensate.

Equation~\eqref{eq:Zi_general} provides the general structural criterion for condensation in terms of the counted microscopic weights. The general theory does not require one to impose a specific asymptotic form on $\Omega_i(w)$. Particular asymptotic forms may be studied as examples, but the condensation criterion itself is simply the finiteness of the critical mean obtained from the partition sum derived by counting.

\section{Instability-Driven Concentration as a Distinct Route}

The capacity-driven transition described above is the primary mechanism of interest in this paper. It should be distinguished from a separate possibility: the absence of a normalizable regular sector altogether.

The failure of the local concavity condition $h_i''(w)<0$ is not, by itself, enough to identify instability-driven concentration. A bounded increase of the curvature ratio can instead lead to a well-defined canonical regular sector with a finite critical capacity. Instability-driven concentration is the stronger case in which the microscopic weight grows so rapidly that
\begin{equation}
    Z_i(\lambda)=\sum_{w=0}^{\infty}\Omega_i(w)e^{-\lambda w}
\end{equation}
diverges for every finite $\lambda$, or equivalently the regular-sector reservoir description cannot be normalized.

In this regime, wealth concentration occurs because the measure is biased toward the boundary of the wealth simplex. Typical fluctuations are no longer small compared with the total wealth, the reservoir expansion loses its regular-sector interpretation, and no finite intensive parameter controls a normalizable marginal distribution.

This is distinct from stable capacity-driven condensation. In the capacity-driven case, the regular sector remains described by a well-defined marginal distribution for $\lambda>\lambda_c$, reaches a finite critical capacity as $\lambda\downarrow\lambda_c$, and then remains pinned at its critical distribution while excess wealth forms the condensate. In the instability-driven case, no such regular critical sector exists. The present work focuses on the former mechanism.

\section{Explicit Examples and Regime Classification}\label{sec:examples}

The following examples illustrate the three regimes identified above. They are not additional assumptions in the general theory. Rather, they are explicit choices of value-wealth geometry used to show how the counting-derived criterion classifies different outcomes.

\subsection{Example I: capacity-driven condensation}

Consider the large-wealth form
\begin{equation}
    V_i'(w) \sim A_i e^{-\mu_i w}(w+w_0)^p,
    \qquad
    A_i>0,\quad \mu_i>0,\quad p>2,\quad w_0>0.
    \label{eq:example_capacity_vp}
\end{equation}
A smooth short-wealth regularization can always be chosen; the condensation criterion depends only on the large-$w$ behavior. If one wants Eq.~\eqref{eq:example_capacity_vp} itself to satisfy $V_i''(w)<0$ for all $w\geq0$, it is sufficient to choose $w_0>p/\mu_i$, since
\begin{equation}
    V_i''(w)
    =A_i e^{-\mu_i w}(w+w_0)^p
    \left(\frac{p}{w+w_0}-\mu_i\right).
\end{equation}
Thus $V_i'(w)>0$ and $V_i''(w)<0$ under this condition.

The counted microscopic weight is
\begin{equation}
    \Omega_i(w)\propto \frac{1}{V_i'(w)}
    \sim A_i^{-1}e^{\mu_i w}(w+w_0)^{-p}.
\end{equation}
The regular-sector partition function has the asymptotic form
\begin{equation}
    Z_i(\lambda)
    \sim A_i^{-1}\sum_{w=0}^{\infty}(w+w_0)^{-p}e^{-(\lambda-\mu_i)w}.
\end{equation}
Therefore
\begin{equation}
    \lambda_{c,i}=\mu_i.
\end{equation}
Using the Lerch transcendent
\begin{equation}
    \Phi(z,s,a)=\sum_{w=0}^{\infty}z^w(w+a)^{-s},
\end{equation}
one may write
\begin{equation}
    Z_i(\lambda)
    \sim A_i^{-1}\Phi\!\left(e^{-(\lambda-\mu_i)},p,w_0\right).
\end{equation}
The mean wealth is
\begin{equation}
    \langle w\rangle_i(\lambda)
    =
    \frac{\Phi\!\left(e^{-(\lambda-\mu_i)},p-1,w_0\right)}
    {\Phi\!\left(e^{-(\lambda-\mu_i)},p,w_0\right)}
    -w_0.
\end{equation}
At criticality,
\begin{equation}
    \langle w\rangle_{i,c}
    =
    \frac{\zeta(p-1,w_0)}{\zeta(p,w_0)}-w_0,
    \label{eq:critical_mean_zeta}
\end{equation}
where $\zeta(s,a)$ is the Hurwitz zeta function. This quantity is finite precisely when $p>2$. Hence the regular sector has finite capacity, and for $W>W_c$ the excess wealth forms a condensate:
\begin{equation}
    W_{\rm cond}=W-W_c.
\end{equation}

This example also illustrates why the local curvature-ratio criterion must be distinguished from the partition-sum criterion. Here
\begin{equation}
    -\frac{V_i''(w)}{V_i'(w)}=\mu_i-\frac{p}{w+w_0},
\end{equation}
which increases toward the finite limiting value $\mu_i$. Therefore this is not a decreasing-curvature-ratio interior-saddle example. It is instead a stable canonical regular-sector example with finite critical capacity.

\subsection{Example II: stable regular sector with no condensation}

Now choose
\begin{equation}
    V_i'(w)=A_i(w+w_0)^{-a},
    \qquad
    A_i>0,\quad a>0,\quad w_0>0.
\end{equation}
Then
\begin{equation}
    V_i'(w)>0,
    \qquad
    V_i''(w)=-aA_i(w+w_0)^{-a-1}<0.
\end{equation}
The microscopic weight is
\begin{equation}
    \Omega_i(w)\propto (w+w_0)^a,
\end{equation}
and the single-agent entropy contribution is
\begin{equation}
    h_i(w)=\ln\Omega_i(w)=a\ln(w+w_0)+\mathrm{const.}
\end{equation}
Thus
\begin{equation}
    h_i''(w)=-\frac{a}{(w+w_0)^2}<0,
\end{equation}
so the local interior entropy geometry is stable.

The partition function is
\begin{equation}
    Z_i(\lambda)\propto \sum_{w=0}^{\infty}(w+w_0)^a e^{-\lambda w}.
\end{equation}
It is finite for $\lambda>0$, so $\lambda_{c,i}=0$. However, as $\lambda\downarrow0$,
\begin{equation}
    \langle w\rangle_i(\lambda)\to\infty.
\end{equation}
Therefore
\begin{equation}
    \langle w\rangle_{i,c}=\infty,
    \qquad
    W_c=\infty.
\end{equation}
There is no finite regular-sector capacity and hence no capacity-driven condensation for any finite total wealth. This is the fully regular case: the regular sector can absorb arbitrarily large wealth by tuning $\lambda$ toward zero.

\subsection{Example III: instability-driven concentration}

Finally choose
\begin{equation}
    V_i'(w)=A_i e^{-b(w+w_0)^2},
    \qquad
    A_i>0,\quad b>0,\quad w_0>0.
\end{equation}
Then
\begin{equation}
    V_i'(w)>0,
    \qquad
    V_i''(w)=-2b(w+w_0)A_i e^{-b(w+w_0)^2}<0.
\end{equation}
Thus the value function is still increasing and concave. However,
\begin{equation}
    \Omega_i(w)\propto e^{b(w+w_0)^2},
\end{equation}
so
\begin{equation}
    h_i(w)=b(w+w_0)^2+\mathrm{const.},
    \qquad
    h_i''(w)=2b>0.
\end{equation}
The entropy contribution is convex, not concave, and an interior stationary point is not an entropy maximum. Equivalently,
\begin{equation}
    -\frac{V_i''(w)}{V_i'(w)}=2b(w+w_0),
\end{equation}
which increases without bound.

The partition function is
\begin{equation}
    Z_i(\lambda)\propto \sum_{w=0}^{\infty}e^{b(w+w_0)^2-\lambda w}.
\end{equation}
This diverges for every finite $\lambda$, because the quadratic growth in the exponent dominates the linear damping term $-\lambda w$. Thus there is no normalizable regular-sector distribution. Wealth concentration is therefore instability-driven rather than capacity-driven.

\subsection{Summary of the three regimes}

\begin{center}
\small
\begin{tabularx}{\textwidth}{@{}p{0.25\textwidth}p{0.27\textwidth}p{0.22\textwidth}X@{}}
\toprule
Case & Example $V_i'(w)$ & Regular sector & Outcome \\
\midrule
Capacity-driven condensation
& $e^{-\mu_i w}(w+w_0)^p$, $p>2$
& finite critical capacity
& condensate for $W>W_c$ \\
No condensation
& $(w+w_0)^{-a}$, $a>0$
& stable, infinite capacity
& no condensate for finite $W$ \\
Instability-driven concentration
& $e^{-b(w+w_0)^2}$
& no normalizable regular sector
& boundary-dominated concentration \\
\bottomrule
\end{tabularx}
\end{center}

\subsection{Consistency of the intensive parameter}

The parameter $\lambda$ appearing in the reservoir expansion is the same intensive parameter that controls the normalizability of the regular-sector distribution. For a fully regular state, the global wealth constraint fixes $\lambda$ through
\begin{equation}
    W=\sum_i\langle w\rangle_i(\lambda).
\end{equation}
If the right-hand side approaches a finite limit as $\lambda\downarrow\lambda_c$, then $\lambda$ is pinned at $\lambda_c$ once $W>W_c$, and the additional wealth forms the condensate. If the right-hand side diverges instead, then $W_c=\infty$ and no capacity-driven condensation occurs.

\section{Open Systems: Variable Wealth and Agent Number}

The preceding analysis has focused on a closed system with fixed total wealth and a fixed number of agents. In many economic environments, however, neither assumption holds strictly. Wealth may be injected or dissipated through growth, credit creation, taxation, or consumption, while the number of participating agents may change through entry, exit, or institutional restructuring. In this section, we show that the structural results derived so far can be extended to such open settings without modifying the underlying logic of microstate counting.

\subsection{Subsystem-environment construction}

We consider a large closed environment characterized by a fixed total wealth $W_{\rm tot}$ and a fixed total number of agents $N_{\rm tot}$. Within this environment, we identify a finite subsystem consisting of $n$ agents holding a total wealth $w$. The subsystem is assumed to be small compared to the environment, but otherwise arbitrary. Because all microstates of the full environment are equally likely, the probability that the subsystem realizes a configuration with total wealth $w$ and agent number $n$ is proportional to the number of compatible microstates of the subsystem and the remaining environment:
\begin{equation}
    P_S(w,n)\propto \Omega_S(w,n)\Omega_R(W_{\rm tot}-w,N_{\rm tot}-n),
\end{equation}
where $\Omega_S$ and $\Omega_R$ denote the microstate counts of the subsystem and the remainder, respectively.

\subsection{Emergent weighting in wealth and agent number}

For a sufficiently large environment, the logarithm of the reservoir multiplicity may be expanded about $(W_{\rm tot},N_{\rm tot})$:
\begin{equation}
    \ln\Omega_R(W_{\rm tot}-w,N_{\rm tot}-n)
    \simeq \ln\Omega_R(W_{\rm tot},N_{\rm tot})-\lambda w-\gamma n,
\end{equation}
where
\begin{equation}
    \lambda\equiv \left.\frac{\partial\ln\Omega_R}{\partial W}\right|_*,
    \qquad
    \gamma\equiv \left.\frac{\partial\ln\Omega_R}{\partial N}\right|_*.
\end{equation}
Higher-order terms are negligible provided $w\ll W_{\rm tot}$ and $n\ll N_{\rm tot}$. Substitution gives
\begin{equation}
    P_S(w,n)\propto \Omega_S(w,n)e^{-\lambda w-\gamma n}.
\end{equation}
Thus, effective exponential weighting in both total wealth and agent number emerges directly from microstate counting, without invoking optimization principles or entropy maximization assumptions at the subsystem level.

\subsection{Agent-level statistics}

Adopting the same coarse-graining as in the closed system, a subsystem macrostate is specified by the wealth vector $\bm{w}=(w_1,\ldots,w_n)$, and the subsystem multiplicity factorizes as
\begin{equation}
    \Omega_S(\bm{w},n)=\prod_{i=1}^{n}\Omega_i(w_i).
\end{equation}
Using the Jacobian postulate, the joint distribution becomes
\begin{equation}
    P(\bm{w},n)\propto \prod_{i=1}^{n}\left[\frac{e^{-\lambda w_i}}{V_i'(w_i)}\right]e^{-\gamma n}.
\end{equation}
Conditional on a fixed number of agents $n$, the wealths follow the same counted regular-sector form,
\begin{equation}
    P_i(w)=\frac{1}{Z_i(\lambda)}\frac{e^{-\lambda w}}{V_i'(w)}.
\end{equation}
The parameter $\gamma$ controls the likelihood of different agent numbers but does not affect the wealth distribution conditional on $n$.

\subsection{Condensation in open systems}

The possibility of wealth condensation in open systems follows the same structural criterion identified earlier. Define the normal-sector mean wealth density
\begin{equation}
    \rho(\lambda)=\frac{\sum_w w\,\Omega(w)e^{-\lambda w}}{\sum_w \Omega(w)e^{-\lambda w}},
\end{equation}
and the corresponding critical density
\begin{equation}
    \rho_c=\lim_{\lambda\downarrow\lambda_c}\rho(\lambda),
\end{equation}
where $\lambda_c$ denotes the lower edge of the normalizable regime. If the subsystem is driven or conditioned to a mean wealth per agent $\rho=w/n$ exceeding $\rho_c$, the regular sector cannot absorb the excess wealth. In this case, a finite fraction of the total wealth accumulates onto a small subset of agents, forming a condensate.

Allowing both wealth and the number of agents to vary leads to a richer interpretation of wealth concentration. Accumulation may occur not only because wealth exceeds the absorptive capacity of existing agents, but also because the creation of new agents is constrained. When demographic accommodation is limited, excess wealth is forced to concentrate even in the absence of instability in the underlying entropy structure.

\section{Weak Time Dependence and Critical Susceptibility}
\label{sec:time_dependence}

We now introduce a minimal time-dependent extension without replacing the counting-based construction by an explicit exchange model. For clarity, the derivation is written for a regular sector with fixed agent number $N$; the same quasistatic logic applies conditionally to an open subsystem when its instantaneous agent number is specified. The central assumption is that external wealth injection, dissipation, or structural changes in $V_i$ occur slowly compared with the equilibration of the regular sector.

\subsection{Weakly driven quasistatic extension}

Introduce a slow time variable
\begin{equation}
    \tau=\epsilon t,
    \qquad
    0<\epsilon\ll 1,
    \label{eq:slow_time}
\end{equation}
where $t$ is physical time. The parameter $\epsilon$ controls the rate of driving, not necessarily its amplitude. We allow
\begin{equation}
    W=W(\tau),
    \qquad
    V_i=V_i(w,\tau),
    \qquad
    \lambda=\lambda(\tau).
    \label{eq:slow_parameters}
\end{equation}
The instantaneous microscopic weight and partition sum are
\begin{equation}
    \Omega_i(w,\tau)
    \propto
    \frac{1}{\partial_w V_i(w,\tau)},
    \qquad
    Z_i(\lambda,\tau)
    =
    \sum_{w=0}^{\infty}\Omega_i(w,\tau)e^{-\lambda w}.
    \label{eq:instantaneous_objects}
\end{equation}
The corresponding quasistatic distribution is
\begin{equation}
    \boxed{
    P_i^{\qs}(w,\tau)
    =
    \frac{\Omega_i(w,\tau)e^{-\lambda(\tau)w}}
    {Z_i(\lambda(\tau),\tau)}.
    }
    \label{eq:quasistatic_distribution}
\end{equation}
This is not a new probabilistic ansatz. It is the original counting-derived marginal evaluated using the instantaneous value--wealth geometry and the instantaneous wealth constraint.

\subsection{Constraint-induced evolution of the intensive parameter}

Before condensation, the common parameter is fixed implicitly by
\begin{equation}
    \boxed{
    W(\tau)
    =
    \sum_i \mean{w}_i(\lambda(\tau),\tau)
    =
    -\sum_i\frac{\partial}{\partial\lambda}
    \ln Z_i(\lambda(\tau),\tau).
    }
    \label{eq:instantaneous_constraint_time}
\end{equation}
Differentiating with respect to $\tau$ gives
\begin{equation}
    \frac{\dd W}{\dd\tau}
    =
    \sum_i
    \left[
    \frac{\partial\mean{w}_i}{\partial\lambda}
    \frac{\dd\lambda}{\dd\tau}
    +
    \left(
    \frac{\partial\mean{w}_i}{\partial\tau}
    \right)_{\lambda}
    \right].
    \label{eq:time_constraint_derivative}
\end{equation}
Because
\begin{equation}
    \frac{\partial\mean{w}_i}{\partial\lambda}
    =-\Var_i(w),
    \label{eq:variance_identity_time}
\end{equation}
we obtain
\begin{equation}
    \boxed{
    \frac{\dd\lambda}{\dd\tau}
    =
    -
    \frac{
    \displaystyle
    \frac{\dd W}{\dd\tau}
    -
    \sum_i
    \left(
    \frac{\partial\mean{w}_i}{\partial\tau}
    \right)_{\lambda}
    }{
    \displaystyle
    \sum_i\Var_i(w)
    }.
    }
    \label{eq:lambda_tau_general}
\end{equation}
Equivalently, in physical time,
\begin{equation}
    \boxed{
    \dot\lambda
    =
    -
    \frac{
    \displaystyle
    \dot W
    -
    \sum_i
    \left(
    \partial_t\mean{w}_i
    \right)_{\lambda}
    }{
    \displaystyle
    \sum_i\Var_i(w)
    }.
    }
    \label{eq:lambda_dot_general}
\end{equation}
If the value--wealth functions are static and only total wealth changes, this reduces to
\begin{equation}
    \boxed{
    \dot\lambda
    =-
    \frac{\dot W}{\sum_i\Var_i(w)}.
    }
    \label{eq:lambda_dot_simple}
\end{equation}
Thus, positive wealth injection drives $\lambda$ downward toward the convergence boundary. Equation~\eqref{eq:lambda_dot_general} is a constraint-induced evolution law: it determines the response of $\lambda$ once $W(t)$ and the drift of $V_i$ have been specified. It is not an autonomous microscopic dynamics and does not itself predict transaction rates or relaxation times.

\subsection{Time-dependent threshold and condensate}

For each sector, define the instantaneous normalizability boundary
\begin{equation}
    \lambda_{c,i}(\tau)
    =
    \inf\left\{
    \lambda:Z_i(\lambda,\tau)<\infty
    \right\},
    \label{eq:lambda_ci_time}
\end{equation}
and the common boundary
\begin{equation}
    \boxed{
    \lambda_c(\tau)=\max_i\lambda_{c,i}(\tau).
    }
    \label{eq:lambda_c_time}
\end{equation}
The instantaneous regular-sector capacity is
\begin{equation}
    \boxed{
    W_c(\tau)
    =
    \sum_i\mean{w}_{i,c}(\tau),
    \qquad
    \mean{w}_{i,c}(\tau)
    =
    \lim_{\lambda\downarrow\lambda_c(\tau)}
    \mean{w}_i(\lambda,\tau),
    }
    \label{eq:Wc_time}
\end{equation}
provided the sum is finite. The quasistatic condensate wealth is therefore
\begin{equation}
    \boxed{
    W_{\rm cond}(\tau)
    =
    \left[W(\tau)-W_c(\tau)\right]_+,
    \qquad
    [x]_+=\max(x,0).
    }
    \label{eq:Wcond_time}
\end{equation}
When $W(\tau)>W_c(\tau)$ and a critical regular sector exists, $\lambda(\tau)$ is pinned at $\lambda_c(\tau)$ and additional wealth is carried by the condensate. Condensation may arise either because $W(\tau)$ increases through a fixed capacity or because structural evolution lowers $W_c(\tau)$ at fixed total wealth. The second route represents time-dependent concentration driven by a deterioration of regular-sector absorptive capacity.

\subsection{Unified first-order response}

The slow-driving parameter $\epsilon$ controls the rate of change. To describe a separate small-amplitude perturbation, introduce $0<\eta\ll1$ and write
\begin{align}
    V_i(w,\tau)
    &=V_i^{(0)}(w)+\eta V_i^{(1)}(w,\tau),
    \label{eq:V_perturbation_time}\\
    W(\tau)
    &=W_0+\eta W_1(\tau),
    \label{eq:W_perturbation_time}\\
    \lambda(\tau)
    &=\lambda_0+\eta\lambda_1(\tau).
    \label{eq:lambda_perturbation_time}
\end{align}
Define
\begin{equation}
    \phi_i(w,\tau)
    =
    \frac{V_i^{(1)\prime}(w,\tau)}{V_i^{(0)\prime}(w)}.
    \label{eq:phi_time}
\end{equation}
Then
\begin{equation}
    \Omega_i(w,\tau)
    =
    \Omega_i^{(0)}(w)
    \left[1-\eta\phi_i(w,\tau)\right]
    +O(\eta^2).
    \label{eq:Omega_first_order_time}
\end{equation}
With the reference distribution
\begin{equation}
    P_i^{(0)}(w)
    =
    \frac{\Omega_i^{(0)}(w)e^{-\lambda_0w}}
    {Z_i^{(0)}(\lambda_0)},
    \label{eq:P0_time}
\end{equation}
the normalized marginal becomes
\begin{equation}
    \boxed{
    P_i(w,\tau)
    =
    P_i^{(0)}(w)
    \left\{
    1-\eta
    \left[
    \phi_i(w,\tau)
    +\lambda_1(\tau)w
    -\mean{\phi_i+\lambda_1w}_{0,i}
    \right]
    \right\}
    +O(\eta^2).
    }
    \label{eq:P_first_order_time}
\end{equation}
The expectation-value subtraction guarantees normalization through first order. The induced change in sectoral mean wealth is
\begin{equation}
    \mean{w}_i-\mean{w}_{0,i}
    =
    -\eta\Cov_{0,i}(w,\phi_i)
    -\eta\lambda_1\Var_{0,i}(w)
    +O(\eta^2).
    \label{eq:mean_first_order_time}
\end{equation}
Imposing the first-order wealth constraint gives the unified correction
\begin{equation}
    \boxed{
    \lambda_1(\tau)
    =
    -
    \frac{
    \displaystyle
    W_1(\tau)
    +
    \sum_i\Cov_{0,i}\!\left(w,\phi_i(w,\tau)\right)
    }{
    \displaystyle
    \sum_i\Var_{0,i}(w)
    }.
    }
    \label{eq:lambda_one_unified}
\end{equation}
A positive wealth perturbation lowers $\lambda$. Structural drift contributes through the covariance between wealth and the fractional change in $V_i'(w)$. The wealth-only and structure-only limits follow by setting $\phi_i=0$ and $W_1=0$, respectively.

\subsection{Wealth susceptibility near the critical boundary}

Define the regular-sector wealth susceptibility
\begin{equation}
    \boxed{
    \chi_W(\lambda)
    \equiv
    -\frac{\partial W}{\partial\lambda}
    =
    \sum_i\Var_i(w).
    }
    \label{eq:wealth_susceptibility}
\end{equation}
Suppose that at least one sector whose individual boundary equals the common boundary,
\begin{equation}
    \mathcal{C}
    =
    \left\{i:\lambda_{c,i}=\lambda_c\right\},
    \label{eq:critical_sector_set}
\end{equation}
has a critical marginal with tail
\begin{equation}
    P_{i,c}(w)\sim A_iw^{-p_i}.
    \label{eq:critical_power_tail}
\end{equation}
The critical mean of that sector is finite for $p_i>2$, whereas its critical variance is finite only for $p_i>3$. Hence a critical sector with
\begin{equation}
    2<p_i<3
    \label{eq:susceptibility_exponent_range}
\end{equation}
contributes a finite condensation capacity but a divergent infinite-system susceptibility. Sectors with $\lambda_{c,i}<\lambda_c$ remain exponentially regularized at the common boundary and do not generate this divergence.

For the shifted critical power law appearing in Example~I,
\begin{equation}
    P_c(w)
    =
    \frac{(w+w_0)^{-p}}{\zeta(p,w_0)},
    \qquad
    w=0,1,2,\ldots,
    \label{eq:shifted_critical_power_law}
\end{equation}
the mean is given by Eq.~\eqref{eq:critical_mean_zeta}. For $p>3$, the variance is
\begin{equation}
    \Var_c(w)
    =
    \frac{\zeta(p-2,w_0)}{\zeta(p,w_0)}
    -
    \left[
    \frac{\zeta(p-1,w_0)}{\zeta(p,w_0)}
    \right]^2.
    \label{eq:critical_variance_zeta}
\end{equation}
For $2<p\leq3$, the defining second-moment series diverges; Eq.~\eqref{eq:critical_variance_zeta} must therefore not be interpreted through analytic continuation of the Hurwitz zeta function.

Let $\delta\lambda=\lambda-\lambda_c>0$. For a critical sector with $2<p<3$, exponential regularization above the boundary gives
\begin{equation}
    \chi_W(\delta\lambda)
    \sim
    C(\delta\lambda)^{p-3}
    =
    \frac{C}{(\delta\lambda)^{3-p}},
    \qquad
    C>0.
    \label{eq:susceptibility_scaling}
\end{equation}
Consequently,
\begin{equation}
    \boxed{
    \frac{\partial\lambda}{\partial W}
    =-
    \frac{1}{\chi_W}
    \longrightarrow0
    \qquad
    \text{as}\quad
    \lambda\downarrow\lambda_c.
    }
    \label{eq:critical_flattening}
\end{equation}
This is a divergent wealth susceptibility, or equivalently a critical flattening of $\lambda(W)$: increasingly large changes in regular-sector wealth correspond to increasingly small changes in the conjugate parameter.

This flattening should not automatically be identified with dynamical critical slowing down. For static $V_i$ and constant wealth injection $\dot W=g>0$, Eqs.~\eqref{eq:lambda_dot_simple} and \eqref{eq:susceptibility_scaling} imply
\begin{equation}
    \dot{\delta\lambda}
    \sim
    -\frac{g}{C}(\delta\lambda)^{3-p}.
    \label{eq:delta_lambda_time}
\end{equation}
For $2<p<3$,
\begin{equation}
    (\delta\lambda)^{p-2}
    =
    (\delta\lambda_0)^{p-2}
    -
    \frac{(p-2)g}{C}t.
    \label{eq:finite_time_boundary}
\end{equation}
The slope $\dot\lambda$ vanishes at the boundary, but the boundary may still be reached in finite time. Genuine critical slowing down requires a divergence of an intrinsic relaxation time, which cannot be inferred from the static counting measure alone.

In a finite economy, an effective upper cutoff $w_{\max}$ regularizes the divergence. For $2<p<3$,
\begin{equation}
    \chi_{W,c}\sim w_{\max}^{\,3-p},
    \label{eq:finite_cutoff_susceptibility}
\end{equation}
while the growth is logarithmic at $p=3$ and remains bounded for $p>3$. The logarithmic boundary case $p=3$ is included explicitly in the information-geometric classification of Sec.~\ref{sec:information_geometry}. Thus an empirically estimated exponent in the interval $2<p<3$ should be interpreted as producing a large, cutoff-sensitive susceptibility rather than a literal infinite variance in a finite dataset.

\subsection{Adiabaticity and finite-relaxation corrections}

Let $\tau_{\rm rel}$ be the intrinsic relaxation time of the regular sector. Sufficient local quasistatic conditions are
\begin{equation}
    \tau_{\rm rel}
    \left|\partial_t\ln\Omega_i(w,t)\right|
    \ll1,
    \label{eq:adiabatic_omega_time}
\end{equation}
and
\begin{equation}
    \tau_{\rm rel}|\dot\lambda|\,\Delta w_i\ll1,
    \label{eq:adiabatic_lambda_time}
\end{equation}
where $\Delta w_i$ is a characteristic distribution width. For static $V_i$, finite $\dot W$, and $2<p<3$,
\begin{equation}
    \Delta w\sim(\delta\lambda)^{-(3-p)/2},
    \qquad
    |\dot\lambda|\sim(\delta\lambda)^{3-p},
\end{equation}
so that
\begin{equation}
    |\dot\lambda|\Delta w
    \sim
    (\delta\lambda)^{(3-p)/2}
    \longrightarrow0.
    \label{eq:adiabatic_product_time}
\end{equation}
If $\tau_{\rm rel}$ remains finite, this quasistatic condition becomes easier to satisfy near the boundary. If instead
\begin{equation}
    \tau_{\rm rel}\sim(\delta\lambda)^{-\gamma},
\end{equation}
then
\begin{equation}
    \tau_{\rm rel}|\dot\lambda|\Delta w
    \sim
    (\delta\lambda)^{(3-p)/2-\gamma}.
    \label{eq:adiabatic_breakdown_time}
\end{equation}
The condition fails asymptotically when $\gamma>(3-p)/2$ and is marginal at equality. Determining $\gamma$ requires a master equation, transition kernel, or another microscopic wealth-exchange model.

A minimal phenomenological lag behind the instantaneous state may be represented by
\begin{equation}
    \frac{\partial P_i(w,t)}{\partial t}
    =
    -\frac{1}{\tau_i}
    \left[
    P_i(w,t)-P_i^{\qs}(w,t)
    \right].
    \label{eq:relaxation_model_time}
\end{equation}
For slow driving,
\begin{equation}
    P_i(w,t)
    \simeq
    P_i^{\qs}(w,t)
    -
    \tau_i\frac{\partial P_i^{\qs}(w,t)}{\partial t}.
    \label{eq:lag_correction_time}
\end{equation}
This correction remains phenomenological. A microscopic theory would require transition rates $K_i(w\mid w',t)$ and a master equation consistent with wealth conservation and, where appropriate, the instantaneous counting-derived stationary measure.

The resulting construction is therefore best described as a weakly driven sequence of counting-derived stationary states. It introduces controlled temporal evolution while preserving the original microstate-counting logic and clearly marks where additional dynamical assumptions become necessary.

\section{Information-Geometric Structure of the Regular Sector}
\label{sec:information_geometry}

The susceptibility introduced in Sec.~\ref{sec:time_dependence} has an exact statistical interpretation. In this section we express that interpretation in the language of information geometry \cite{amari2016} and connect it to the condensation classification. To distinguish the entropy of an aggregate constrained state from the macrostate entropy $S(\bm w)$ introduced earlier, we denote the former by $\mathcal{S}(W)$.

\subsection{Joint Fisher information and Legendre duality}

For the canonical regular sector, define the joint product measure and joint partition function
\begin{equation}
    P_\lambda(\bm w)=\prod_{i=1}^{N}P_{i,\lambda}(w_i),
    \qquad
    Z(\lambda)=\prod_{i=1}^{N}Z_i(\lambda),
\end{equation}
with
\begin{equation}
    F(\lambda)\equiv\ln Z(\lambda)=\sum_i\ln Z_i(\lambda),
    \qquad
    W(\lambda)=-F'(\lambda).
    \label{eq:infogeo_free_energy}
\end{equation}
The aggregate constrained entropy is the Legendre--Fenchel transform
\begin{equation}
    \mathcal{S}(W)
    =
    \inf_{\lambda\geq\lambda_c}
    \left[\lambda W+F(\lambda)\right].
    \label{eq:infogeo_legendre}
\end{equation}
On the regular interior branch, $\mathcal{S}'(W)=\lambda$.

The score of one marginal is
\begin{equation}
    \partial_\lambda\ln P_{i,\lambda}(w)
    =-
    \left(w-\mean{w}_i\right),
\end{equation}
so the single-agent Fisher information is $I_i(\lambda)=\Var_i(w)$. The susceptibility is a sum over agents and therefore equals the Fisher information of the joint product family:
\begin{equation}
    I_{\rm joint}(\lambda)
    =
    \Var_\lambda\!\left(\sum_iw_i\right)
    =
    \sum_i\Var_i(w)
    =
    \chi_W(\lambda),
    \label{eq:joint_fisher_information}
\end{equation}
where canonical independence has been used. Thus the response coefficient in Eq.~\eqref{eq:wealth_susceptibility} is exactly the Fisher information of the joint wealth family.

In the natural coordinate $\lambda$, the Fisher--Rao metric is
\begin{equation}
    g_{\lambda\lambda}
    =
    I_{\rm joint}(\lambda)
    =
    \chi_W(\lambda)
    =
    F''(\lambda).
    \label{eq:metric_lambda}
\end{equation}
Since
\begin{equation}
    \frac{\dd W}{\dd\lambda}=-\chi_W,
    \qquad
    \frac{\dd\lambda}{\dd W}=-\frac{1}{\chi_W},
    \qquad
    \mathcal{S}''(W)=-\frac{1}{\chi_W},
    \label{eq:legendre_metric_derivatives}
\end{equation}
its component in the dual expectation coordinate is
\begin{equation}
    g_{WW}
    =
    g_{\lambda\lambda}
    \left(\frac{\dd\lambda}{\dd W}\right)^2
    =
    \frac{1}{\chi_W}
    =
    -\mathcal{S}''(W).
\end{equation}
Equivalently,
\begin{equation}
    \boxed{
    \dd s^2
    =
    \chi_W\,\dd\lambda^2
    =
    \chi_W^{-1}\,\dd W^2
    =
    -\mathcal{S}''(W)\,\dd W^2
    },
    \qquad
    g^{WW}=\chi_W=-\frac{1}{\mathcal{S}''(W)}.
    \label{eq:dual_metric_identity}
\end{equation}

\begin{proposition}[Dual metric identity]
On the regular sector, the wealth susceptibility is the Fisher metric component in the natural coordinate, $g_{\lambda\lambda}=\chi_W=F''(\lambda)$, whereas the negative constrained-entropy curvature is the component in the dual expectation coordinate, $g_{WW}=-\mathcal{S}''(W)=\chi_W^{-1}$. These are two coordinate representations of the same one-dimensional metric.
\end{proposition}

Every one-dimensional Riemannian manifold has identically vanishing intrinsic Riemann curvature. A divergence of $g_{\lambda\lambda}$ is therefore not an intrinsic curvature singularity; it is a degeneration of the Legendre coordinate transformation. For $2<p<3$, Eq.~\eqref{eq:susceptibility_scaling} gives
\begin{equation}
    d_{\FR}
    \sim
    \int_0^\epsilon
    \sqrt{\chi_W(\delta\lambda)}\,\dd(\delta\lambda)
    \sim
    \epsilon^{(p-1)/2}
    <\infty.
    \label{eq:fisher_distance_boundary}
\end{equation}
At $p=3$ the divergence is logarithmic and the corresponding distance is also finite, while for $p>3$ the metric remains finite. Hence the finite-capacity boundary lies at finite Fisher--Rao distance throughout $p>2$.

\subsection{Condensation and the information-geometric trichotomy}

The capacity criterion depends on the first moment, whereas the Fisher metric depends on the second. For a critical sector with $P_{i,c}(w)\sim A_iw^{-p}$, this produces the classification in Table~\ref{tab:infogeo_trichotomy}.

\begin{table}[ht]
\centering
\caption{Capacity and information-geometric behavior for a critical power-law sector.}
\label{tab:infogeo_trichotomy}
\renewcommand{\arraystretch}{1.2}
\begin{tabularx}{\textwidth}{cXX}
\toprule
Tail exponent & Regular-sector capacity & Fisher/susceptibility signature \\
\midrule
$p\leq2$ & $W_c=\infty$; no finite-capacity condensation & No finite condensation boundary \\
$2<p<3$ & $W_c<\infty$ & $\chi_W\sim C(\delta\lambda)^{p-3}$ diverges algebraically \\
$p=3$ & $W_c<\infty$ & $\chi_W\sim C\ln(1/\delta\lambda)$ diverges logarithmically \\
$p>3$ & $W_c<\infty$ & $\chi_W\to\chi_{W,c}<\infty$ \\
\bottomrule
\end{tabularx}
\end{table}

If several sectors belong to the critical set $\mathcal{C}$ of Eq.~\eqref{eq:critical_sector_set}, the most singular member controls the leading divergence of the sum in $\chi_W$. Sectors outside $\mathcal{C}$ remain exponentially regularized at the common boundary.

For every finite-capacity case $p>2$, the map $W(\lambda)=-F'(\lambda)$ terminates at $W_c$. When $W>W_c$, no normalizable tilt $\lambda>\lambda_c$ realizes the imposed total wealth: the regular background remains pinned at $\lambda_c$ and the excess $W-W_c$ occupies the condensate. This is a loss of global equivalence between the canonical parameterization and the constrained description, although local and bulk observables can remain equivalent to those of the critical background \cite{grosskinsky2003,evanswaclaw2014}. At leading extensive order, the Legendre--Fenchel continuation is
\begin{equation}
    \mathcal{S}(W)
    =
    \mathcal{S}(W_c)
    +
    \lambda_c(W-W_c),
    \qquad W>W_c.
    \label{eq:affine_entropy_continuation}
\end{equation}
This is a supporting-line continuation forced by saturation at the boundary of the admissible $\lambda$-domain, not a Maxwell construction between two competing thermodynamic branches. Finite systems acquire subextensive condensate and cutoff corrections.

The central distinction is therefore:
\begin{equation}
    \boxed{
    \text{finite-capacity condensation: }p>2,
    \qquad
    \text{Fisher-metric divergence: }2<p\leq3.
    }
\end{equation}
For $2<p\leq3$, $g_{\lambda\lambda}\to\infty$ and $g_{WW}\to0$. For $p>3$, both components remain finite at the endpoint; condensation still occurs because the admissible canonical domain ends at $\lambda_c$, not because the local metric diverges.

\subsection{Qualified Cram\'er--Rao duality}

Let $n_{\rm s}$ denote the number of statistically independent realizations of the complete joint wealth vector $\bm w=(w_1,\ldots,w_N)$. One realization carries Fisher information $\chi_W$, so an unbiased estimator in a correctly specified one-parameter family and at an interior point $\lambda>\lambda_c$ obeys
\begin{equation}
    \Var(\widehat\lambda)
    \geq
    \frac{1}{n_{\rm s}\chi_W(\lambda)},
    \qquad
    \Var(\widehat W)
    \geq
    \frac{\chi_W(\lambda)}{n_{\rm s}},
    \label{eq:cramer_rao_duality}
\end{equation}
where $I_W=I_\lambda(\dd\lambda/\dd W)^2=1/\chi_W$. For a single observed joint economy, $n_{\rm s}=1$. In a homogeneous i.i.d. formulation one may instead begin with the single-agent information and obtain $I_N=N\Var(w)$ from $N$ agents; the two sampling conventions must not be combined.

\begin{corollary}[Estimation duality]
In the divergent-susceptibility regime $2<p\leq3$, approach to the boundary sharpens the local Fisher lower bound for the natural parameter $\lambda$ and worsens the corresponding bound for the expectation parameter $W$. For $p>3$, both bounds approach finite nonzero limits. Condensation itself therefore does not universally imply increasing precision for $\lambda$.
\end{corollary}

Equation~\eqref{eq:cramer_rao_duality} is an interior regular-model statement, not a claim about inference exactly at $\lambda_c$. Boundary likelihoods can exhibit non-Gaussian limits and nonstandard convergence rates \cite{brazzale2024}. Empirical wealth data additionally involve heterogeneous $\Omega_i$, dependence, censoring, unknown $p$ and $\lambda_c$, and finite samples that may omit the rare extremes driving the formal divergence. Consequently, whether the increasing local information yields narrower finite-sample confidence intervals after nuisance parameters are co-estimated is a separate nonregular inference problem.

\subsection{Open-system mixed-response relation}

The open-system construction introduces the conjugate pairs $(\lambda,W)$ and $(\gamma,N)$ through the same reservoir entropy,
\begin{equation}
    \lambda=\frac{\partial\ln\Omega_R}{\partial W},
    \qquad
    \gamma=\frac{\partial\ln\Omega_R}{\partial N}.
\end{equation}
Provided $\ln\Omega_R(W,N)$ is twice differentiable, $N$ is treated in a continuum or thermodynamic approximation, and sign conventions are used consistently, equality of mixed partial derivatives gives
\begin{equation}
    \boxed{
    \left.\frac{\partial\lambda}{\partial N}\right|_W
    =
    \left.\frac{\partial\gamma}{\partial W}\right|_N
    =
    \frac{\partial^2\ln\Omega_R}{\partial W\,\partial N}
    }.
    \label{eq:open_system_reciprocity}
\end{equation}
This is a consistency relation of the model: the response of the inequality parameter to admitting agents equals the response of the agent-admission parameter to injecting wealth. An empirical test would require an operational estimator for $\gamma$ and controlled partial derivatives in both aggregate wealth and population.

\subsection{Geometric conjectures and dynamical closure}

The relation $\lambda=\partial_W\mathcal{S}$ can be phrased on the thermodynamic contact manifold $(\mathcal{S},W,\lambda)$ with contact one-form
\begin{equation}
    \vartheta=\dd\mathcal{S}-\lambda\,\dd W.
\end{equation}
The equilibrium family $\{\mathcal{S}=\mathcal{S}(W),\lambda=\mathcal{S}'(W)\}$ is a Legendrian submanifold satisfying $\vartheta=0$, and its projection onto $(W,\lambda)$ is a Lagrangian graph. An effective Hamiltonian can be chosen to have this graph as a zero level set, but such a choice is non-unique and has no dynamical content unless derived from a stochastic generator.

For $2<p<3$, integration of $\dd W/\dd\lambda=-\chi_W$ gives
\begin{equation}
    W_c-W\sim A(\delta\lambda)^{p-2},
    \qquad
    \delta\lambda\sim B(W_c-W)^{1/(p-2)}.
\end{equation}
At $p=5/2$ this becomes $\lambda-\lambda_c\sim B(W_c-W)^2$, and the concavity of $\mathcal{S}$ gives
\begin{equation}
    \mathcal{S}(W)
    =
    \mathcal{S}_c
    +
    \lambda_c(W-W_c)
    -
    \widetilde C(W_c-W)^3
    +\cdots,
    \qquad
    \widetilde C>0.
    \label{eq:cubic_normal_form}
\end{equation}
The cubic correction is compatible with an $A_2$-type normal form \cite{ehlers2000}. A genuine Airy finite-size law would have to emerge from the inverse-Laplace representation
\begin{equation}
    \Omega_N(W)
    =
    \frac{1}{2\pi i}
    \int_{\mathcal B}\dd\lambda\,
    \exp\!\left[\lambda W+F_N(\lambda)\right]
    \label{eq:inverse_laplace_airylaw}
\end{equation}
through an appropriate saddle coalescence or saddle--endpoint degeneracy. This remains a conjectural direction.

Finally, a zero-range hop rate $u(w)=\Omega(w-1)/\Omega(w)$ reproduces a factorized stationary weight proportional to $\Omega(w)$. Internal hopping conserves total wealth and therefore cannot by itself relax the aggregate $W$. A dynamical Hamiltonian with distinct relaxation and activation branches requires an explicit open birth--death generator, including injection and removal rates, system--reservoir coupling, large-deviation scaling, and a WKB momentum convention. Different stochastic dynamics can share the same stationary measure while exhibiting different condensation kinetics \cite{evanswaclaw2014}; the stationary counting weight therefore does not determine a relaxation time.

\section{Limitations and Scope}

The framework developed in this work is intentionally structural and minimalist, and several limitations should be kept in mind. First, the time-dependent construction is quasistatic rather than microscopic. It determines how the counting-derived regular state and the conjugate parameter respond to externally specified changes in total wealth or value--wealth geometry, but it does not derive those changes from transaction rules. Convergence times, path dependence, and genuine dynamical critical slowing down require an explicit exchange model or transition kernel.

Second, agents are treated as passive carriers of wealth, distinguished only by their value-wealth convertibility functions. Strategic behavior, learning, negotiation, institutional design, and explicit exchange mechanisms are not modeled. While this abstraction allows structural effects to be isolated cleanly, it also means that the theory does not capture feedback between agent behavior and the underlying configuration space.

Third, the value--wealth convertibility function is taken as an exogenous structural input. The weak-time-dependence section permits this function to drift slowly, but does not endogenize that drift or its feedback with wealth allocation. Fourth, the notion of wealth employed in the model is deliberately coarse-grained. Wealth quanta are treated as homogeneous and perfectly transferable, and distinctions between different asset classes, liquidity constraints, or non-monetary resources are neglected.

The information-geometric conclusions are likewise statements about the regular canonical family. Because the state manifold is one-dimensional, it has no intrinsic Riemann-curvature singularity; a divergent $g_{\lambda\lambda}$ reflects the degeneration of the Legendre coordinate map. The Cram\'er--Rao bounds apply only to correctly specified interior models and do not, by themselves, establish improved empirical estimation at a nonregular boundary. The contact-geometric description, the proposed cubic normal form at $p=5/2$, and any Airy finite-size scaling remain interpretive or conjectural until an appropriate finite-system inversion problem is derived.

Finally, the condensation criterion is intentionally conditional. The theory does not claim that diminishing marginal convertibility automatically produces condensation. Rather, it shows how to decide whether stable capacity-driven condensation occurs once the microscopic weights have been obtained from the value-wealth geometry. This conditional form preserves the counting-based origin of the partition sum while avoiding additional assumptions about the asymptotic tail of the weights.

\section{Conclusion and Outlook}

We have developed a structural framework for wealth allocation based on unbiased counting of admissible configurations in a value--wealth space. Wealth is treated as an extensive quantity distributed across agents, while productive structure enters through an agent-dependent value--wealth convertibility function. Entropy is defined directly from microstate counting, and the corresponding microscopic weights determine the regular-sector partition sum through a reservoir construction.

The main result is a stable capacity-driven criterion for wealth condensation. The partition function
\begin{equation}
    Z_i(\lambda)=\sum_{w=0}^{\infty}\Omega_i(w)e^{-\lambda w}
\end{equation}
is not assumed independently; it emerges from the microcanonical marginal. The regular sector has a finite absorption capacity only if the critical mean
\begin{equation}
    \lim_{\lambda\downarrow\lambda_c}
    \left[-\frac{\partial}{\partial\lambda}\ln Z_i(\lambda)\right]
\end{equation}
is finite. When the total wealth exceeds the resulting capacity, the regular sector remains pinned at its critical distribution and the excess wealth forms a condensate.

This formulation separates a fully regular regime with infinite capacity, a capacity-driven regime with finite critical capacity, and an instability-driven regime in which no normalizable regular sector exists. The open-system extension shows how weighting in both wealth and agent number emerges from a larger reservoir, while the weak-time-dependence construction yields an instantaneous threshold, a constraint-induced evolution law for $\lambda(t)$, and a unified response to changes in total wealth and value--wealth geometry. These temporal results describe a slowly driven sequence of stationary states rather than a microscopic exchange dynamics.

The information-geometric extension identifies the wealth susceptibility as the Fisher information of the joint canonical family,
\begin{equation}
    \chi_W=I_{\rm joint}(\lambda)=\sum_i\Var_i(w),
\end{equation}
and Legendre duality gives
\begin{equation}
    \dd s^2
    =
    \chi_W\dd\lambda^2
    =
    \chi_W^{-1}\dd W^2
    =
    -\mathcal{S}''(W)\dd W^2.
\end{equation}
This sharpens the critical classification. A finite condensation capacity requires $p>2$, but a divergent Fisher metric requires the stronger infinite-variance condition $2<p\leq3$; for $p>3$, condensation occurs with finite susceptibility. Thus grand-canonical saturation is the universal mechanism in the finite-capacity sector, whereas metric divergence is a property of only part of that sector. The boundary remains at finite Fisher--Rao distance, and the open-system variables satisfy the mixed-response relation in Eq.~\eqref{eq:open_system_reciprocity}.

Future work may derive microscopic transition dynamics consistent with the counting-based stationary measure, identify when the $p=5/2$ cubic correction produces a genuine Airy finite-size law, endogenize the evolution of value--wealth convertibility, and test the susceptibility, inference duality, and mixed open-system response against empirical or agent-based data. More broadly, the results suggest that configuration-space and information-geometric methods provide complementary ways to understand stable inequality as an emergent property of economic structure.

\section*{Acknowledgments}
The author utilized OpenAI's ChatGPT to assist with grammar refinement and language clarity during the preparation of this manuscript. All conceptual and mathematical content was solely developed by the author.

\end{document}